\documentclass[journal]{IEEEtran}
\usepackage{cite}
\usepackage{amsmath}
\usepackage{amscd}
\usepackage{amssymb}
\usepackage{graphicx}
\usepackage{algorithmic}
\usepackage{url}
\usepackage{caption}
\usepackage{graphicx}
\usepackage{tabularx}
\usepackage{float}
\usepackage{subfig}
\usepackage{mathrsfs}
\usepackage{multirow}
\usepackage{makecell}
\usepackage{diagbox}
\usepackage{amssymb}
\usepackage{algorithm}
\usepackage{algorithmic}
\usepackage{chemarrow}
\usepackage{extarrows}

\begin{document}

\title{A Restrained Paillier Cryptosystem and Its Applications for Access Control of Common Secret}

\author{Xiaojuan~Dong,
        Weiming~Zhang,
        Mohsin~Shah,
        Bei~Wang,
        Nenghai~Yu

\thanks{All the authors are with CAS Key Laboratory of Electro-magnetic Space Information, University of Science and Technology of China, Hefei 230026, China.

Corresponding author: Weiming Zhang (email:zhangwm@ustc.edu.cn).

This work was supported in part by the Natural Science Foundation of China under Grant U1636201 and 61572452, by Anhui Initiative in Quantum Information Technologies under Grant AHY150400.
} 
 }

\maketitle
\begin{abstract}
The modified Paillier cryptosystem  has become extremely popular and applied in many fields, owning to its additive homomorphism. This cryptosystem provides weak private keys and a strong private key. A weak private key only can decrypt ciphertexts under the corresponding public key. The strong private key can decrypt all ciphertexts even under different public keys. When the modified Paillier cryptosystem is applied in a system, the member, often the system administrator, has the strong private key and can decrypt all ciphertexts. If this system administrator is attacked or compromised, the security of the application system absolutely break down. Thus, it is important to stop the decryption of the strong private key. To address this issue, we propose an restrained version of the modified Paillier cryptosystem (Restrained-Paillier), by endowing the multiplicative homomorphism. We perform the additive encryption on the multiplicative ciphertext and generate the mixed ciphertext, which can not be decrypted by the strong private key. Based on this Restrained-Paillier, we develop two applications. Firstly, we realize access control of common secret of two owners. In  our scheme, only one owner cannot access secret. Secondly, we present three protocols for identity distribution and key management, identity authentication and private key recovery. Security analysis shows that the Restrained-Paillier cryptosystem can resist the chosen plaintext attack. The experimental results illustrate the utility and efficiency of the proposed protocols.
\end{abstract}

\begin{IEEEkeywords}
 Homomorphic Encryption; Privacy Preserving; Access Control; Identity Certificates.
\end{IEEEkeywords}

\section{Introduction}
\IEEEPARstart{E}{ncryption} technology is commonly used to protect confidentiality of data. There are two kinds of encryption technology, one kind is the symmetric encryption like DES \cite{DES}, and the other kind is the public key encryption like RSA \cite{Rivest}. However, symmetric encryption  does not support mathematical operations on encrypted data. In order to perform useful operations on encrypted data, we would decrypt the encrypted data, perform the required operations, and then encrypt the calculation result, which is inconvenient. Fortunately, homomorphic encryption, a kind of public key encryption, allows mathematical operations on encrypted data without exposing private information.

Now, homomorphic encryption is popular in many fields. Fully homomorphic encryption (FHE) allows arbitrary computations over ciphertexts. But, FHE has high time and space complexity so it is not yet efficient enough to be implemented on real systems \cite{Morris}.  Partially multiplicative homomorphic encryption schemes, such as RSA cryptosystem \cite{Rivest}, allow multiplication of plaintext data in the ciphertext domain. Partially additive homomorphic encryption schemes support some additive calculations over ciphertexts. The Paillier cryptosystem \cite{Paillier} is an additive homomorphic encryption scheme.
In the Paillier cryptosystem, a module $N$ is used for only a user, accompanied by a decryption key $\lambda$. Cramer and Shoup \cite{Cramer} proposed a variant of the Paillier cryptosystem, where a module $N$ can be used for a group. Members of the group can communicate and perform multi-party security calculation in a common module $N$. Each member of the group has a decryption key $\theta$, while the varied Paillier cryptosystem has the inherent decryption key $\lambda$. $\theta$ is called as the weak decryption key as it only can decrypt its corresponding owner's ciphertext. $\lambda$ is called as the strong key since it can decrypt all ciphertext. Bresson et al. in \cite{Bresson} simplified the scheme in \cite{Cramer} and their cryptosystem works in the cyclic group of quadratic residues modulo $N^2$.

Many researchers have applied the modified Paillier in \cite{Bresson} cryptosystem to their schemes, especially with multiple parties involved. Multiple parties can perform ciphertext calculation, ciphertext search and so on, while protecting privacy data. For instance, Mohanty  et al. in \cite{Mohanty} proposed a modified Paillier cryptosystem-based image processing scheme, where a image outsourcer, a cloud server and an image user were involved. The cloud server in \cite{Mohanty} can perform scaling and cropping operations over encrypted images with the help of the image outsourcer and the image user. Ayday et al. in \cite{Ayday} introduced a privacy-preserving disease susceptibility test. The the system administrator, patients and the pharmaceutical company were involved. The genomic privacy of a patient were preserved from the pharmaceutical company, via the encryption under the patient's public key. However, the system administrator, who had the strong key, may be compromised to the pharmaceutical company for benefits. As a result, the pharmaceutical company could obtain the all patients' genomic data.

The strong key leakage is a threat to the modified Paillier cryptosystem. To solve the issue, Liu  et al. in \cite{Liu2016multiplekeys} split the strong private key into different shares. One share was send to a cloud platform (CP) and the other share was sent to a computation service provider (CSP). Only one split key could not allow decryption. The authors in \cite{Liu2016multiplekeys} did not discuss that the CP and the CSP conspired in order to decrypt user's data. Moreover, they assumed that the Key Generation Center (KGC) was a trusted party so that the KGC could be tasked with the distribution and management of private keys in the system. Nevertheless, this assumption is unreasonable since the MA may reveal private keys for benefits or being invaded by skilled attackers.

The exposure of strong key  can destroy the security of the system. However, the strong key $\lambda$ can not be  abandoned because $\lambda$ is an inherent part of the modified Paillier cryptosystem. In this paper, we propose to restrain the decryption ability of the strong key. Based on this idea, we propose an restrained version of the modified Paillier cryptosystem (Restrained-Paillier). In the Restrained-Paillier system, the multiplicative ciphertext is added. We perform the additive encryption on the multiplicative ciphertext and generate the mixed ciphertext. The strong private key $\lambda$ is used to decrypt the mixed ciphertext, obtaining the multiplicative ciphertext. But the $\lambda$ can not decrypt the multiplicative ciphertext to get the plaintext data. The multiplicative ciphertext only can be decrypted by its corresponding weak key $\theta$.
Hence, a party has the strong key $\lambda$ and still fail to decrypt the mixed  ciphertext data. We regard the KGC as a semi-trusted party. The KGC can manage both public and private keys, but can not obtain private keys. Specifically, the major contribution of this paper can be summarized below.
\begin{itemize}
\item{First, the  Restrained-Paillier cryptosystem supports mixed ciphertexts, which are generated by executing the additive encryption on the multiplicative ciphertexts.  The strong key decrypts the mixed ciphertexts, only getting the multiplicative ciphertexts but not the plaintext data. }
\item{Second, we use the property of mixed ciphertext to realize access control to common secrets belonging to two owner. No one can obtain the the secret by oneself. Besides, a owner can control the other owner's information obtained form their common secret.}
\item{Third, based on the Restrained-Paillier cryptosystem, we present three associated protocols for achieving identity distribution and key management, identity authentication, and private key recovery. The key management protocol can prevent the private key exposure. If some one lost his/her private key, the private key recovery protocol can recover the key. }
\item{Fourth, we give security analysis that the Restrained-Paillier cryptosystem can resist the chosen plaintext attack. The proposed protocols can achieve security requirements.}
\item{Fifth, we analyze the communication and computation overheads of the Restrained-Paillier cryptosystem and protocols. We  examine these overheads by building a simulator in Java to demonstrate the utility of our proposals.}
\end{itemize}
\subsection{Organization}
The rest of this paper is organized as follows. In Section II, we describe the related preliminaries and the relevant background materials. In Section III, we present the Restrained-Paillier cryptosystem. In Section IV, we present four protocols, three of which is about key key management and Identity certificate, the last one is about access control to common secrets. In Section V, we give the security analysis of our proposals. In Section VI, performance of simulator is evaluated.  Section VII concludes this paper.
\section{Preliminary}
In this section, we review the cryptographic primitives involved, and the system model, the security requirements of the proposed protocols in this paper.
\subsection{Cryptographic Primitives}
In this section, we introduce typical properties of partially homomorphic cryptosystems and then review the modified Paillier cryptosystem in \cite{Bresson} and \cite{Cramer}. For the sake of brevity and readability, Table \ref{table_one} lists some notations used for the rest of the paper.
\begin{table}[!thbp]
\caption{Definitions and Notations in the Cryptosystems}
\label{table_one}
\centering
\begin{tabular}{|c|c|}
\hline
Symbols & Definition\\
\hline
$|\cdot|$ & Bit length\\
\hline
$pk$, $sk$ & Partially homomorphic public \& private key\\
\hline
$E^+_{pk}(\cdot)$ & Additive encryption algorithm with public key\\
\hline
$E^\times_{pk}(\cdot)$ & Multiplicative encryption algorithm with public key\\
\hline
$lcm(x, y)$ & Lowest common multiple  between \emph{x} and \emph{y}\\
\hline
$gcd(x, y)$ & Greatest common divisor between \emph{x} and \emph{y}\\
\hline
$H(\cdot)$ & Hash Function\\
\hline
\end{tabular}
\end{table}
\subsubsection{Partially Homomorphic Cryptosystems}
Two additive ciphertexts $E^+_{pk}(m_1)$ and $E^+_{pk}(m_2)$  accord with the following property:

$E^+_{pk}(m_1)\times E^+_{pk}(m_2) = E^+_{pk}(m_1 + m_2)$;

Two multiplicative ciphertexts like $E^\times_{pk}(m_1)$ and $E^\times_{pk}(m_2)$ accord with the following property:

$E^\times_{pk}(m_1) \times E^\times_{pk}(m_2)=E^\times_{pk}(m_1 \times m_2)$.

\begin{figure}[htp]
\centering
\includegraphics[scale=0.5]{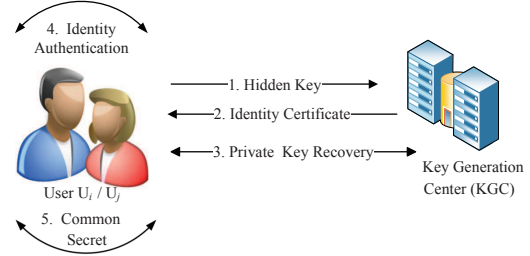}%
\caption{System Model.}
\label{fig_Interact}
\end{figure}
\subsubsection{Paillier Cryptosystem}
The pubic key is $(N, g)$ with $N|{ord_{N^2}}(g)$, where $N = pq$ and $p$ and $q$ are two prime numbers. ${ord_{N^2}}(g)$ is the order of $g$. The private key $\lambda = lcm(p-1, q-1)$. To encrypt a message $m \in \mathbb{Z}_{N}$, select a random $r \in \mathbb{Z}_{N}$ and compute : $c = g^{m+rN}\mod N^2$. The message can be recovered as: $m = L(c^\lambda)/L(g^\lambda)\mod N$, where $L(u) = \frac{u-1\mod N^2}{N}$, for all $u \in \{u<N^2 | u = 1\mod N\}$.

\subsubsection{Modified Paillier Cryptosystem}
The pubic key is $(N, g, h = g^\theta\mod N^2)$ with ${ord_{N^2}}(g)=\lambda =2p'q'$. $\lambda$ is the factorization of $N = pq$ (where $p = 2p' + 1$, $q= 2p' + 1$ are safe primes), and the weak key is $\theta \in [1, {N^2}/{2}]$. Such a $g$ can be easily found by selecting a random $a \in \mathbb{Z}^*_{N^2}$ and calculating $g = -a^{2N}\mod N^2$. To encrypt a message $m \in \mathbb{Z}_{N}$, select a random $r \in [1, N/4]$ and compute : $T_1 = g^r\mod N^2 , T_2 = h^r(1 + mN)\mod N^2$.

Using the weak key $\theta$, the message can be recovered as: $m = L(T_2/{T_1^\theta}\mod N^2)$. If $(p, q)$ is known, then $\lambda=2p'q'$ can be calculated. $T_2^{\lambda} = g^{\lambda (N)xr}(1+m\lambda N) = (1 + m\lambda N)$. Thus, given that $gcd(\lambda, N) = 1$, $m$ can be recovered as: $m = L(T_2^{\lambda}){\lambda}^{-1}\mod N$.

The difference between the Paillier cryptosystem in \cite{Paillier} and the modified Paillier cryptosystem in \cite{Bresson} is the number of users. The Paillier cryptosystem can only be used for one user, who has the public-private key pair $(N, g, \lambda)$. The modified Paillier cryptosystem can be used for a group of users, or an organization. The member of the organization has the public and weak private key pair $(N, g^{\theta}, \theta)$ and only can decrypt the ciphertext under $g^{\theta}$, while the organization manager has the public and strong private key pair $(N, g, \lambda)$ and can decrypt any ciphertext.

\subsection{System Model \& Security Requirement}
\subsubsection{System Model}
The system model consists of three participants: a KGC, a User $U_i$ and a User $U_j$, as illustrated in Fig. \ref{fig_Interact}.

\emph{KGC:} The semi-trusted KGC is tasked with the key management and the distribution of identity certificates in the system. The KGC is curious-but-honest, which strictly follows the protocols, but also interested to learn private keys belonging to users.

\emph{$U_i$:} The $U_i$ registers in the system, sends his/hert hidden key to the KGC and asks the KGC for his/her identity certificate. When the $U_i$'s private key is lost, $U_i$'s can recover his/her key with the the cooperation of the KGC.

\emph{$U_j$:} The $U_j$, a registered user, authenticates the $U_i$'s identity before their communication. There is a common secret between $U_i$ and $U_j$. But no one can open the secret by oneself. Only with the permission of $U_i$, $U_j$ can get secrets and vice versa.
\subsubsection{Design Goals}
To efficiently support key protection, identity authentication and access control of common secret, our  work is designed to achieve the following goals:
\begin{itemize}
\item{\textbf{Access Control of Common Secret}. A common secret belongs to two users. One user cannot open the secret without the permission of his/her companion.}
\item{\textbf{Protection of Private Keys}. To present the KGC knows participants' private keys, private keys are unavailable to the KGC. }
\item{\textbf{Correctness of Identity Certificates}. To guarantee that when a user sends his/her true identity certificate to the other user, the identity certificate can pass the verification of the other user.}
\item{\textbf{Soundness of Identity Certificates}. Assure that corrupted identity certificates cannot pass verification.}
\item{\textbf{Unforgeability of Identity Certificates}. A passive adversary has public keys, but cannot forge a certificate of his chosen identity.}
\item{\textbf{Existential Unforgeability Against Adaptive Chosen Messages Attacks}. An active adversary can access a certificate oracle, which can generate legitimate identity certificates. The adversary's goal is to generate a legitimate identity certificate of an identity, which cannot has been asked the certificate oracle before. Thus, this active adversary win. To stop an active adversary, our proposals are under existential unforgeability. }
\item{\textbf{Recovery of Lost Private Keys}. When a user loses his/her private key, the key can be restored.}
\end{itemize}

\section{Restrained-Paillier Cryptosystem}
The strong  key $\lambda$ of the modified Paillier system can decrypt all ciphertexts. This threatens the security of this system since $\lambda$ may be compromised. In order to address this issue, we endow a multiplicative homomorphic encryption to the modified Paillier, yielding mixed ciphertexts. The mixed ciphertexts are obtained by additively encrypting the multiplicative ciphertexts. $\lambda$ cannot disclose the mixed ciphertexts. The details of the Restrained-Paillier cryptosystem are described as follows.

\textbf{Key Generation (KeyGen):} The public key is $(N, g, h = g^\theta\mod N)$ with the base $g$ (where $p = 2p' + 1, q = 2q' +1$ are safe primes). The strong key is $\lambda$, and the weak private key $\theta \in [1, N^2/2]$. Such a $g$ can be obtained by select a random $a \in {\mathbb{Z}}^*_{N^2}$ but $a \neq 1\mod N$ and computing $g = -a^{2N}\mod N$. The base $g$  meets $g^\lambda\mod N = 1$ and $g^{N\lambda}\mod N^2 = 1$, proved in Appendix A.  The user $U_i$s' a pair of public and private key is denoted as $({pk}_i,{sk}_i)$.  The user $U_j$'s a pair of keys is denoted as $({pk}_j,{sk}_j)$.

\textbf{Additive Encryption (AddEnc):} Given a message $m \in {\mathbb{Z}}_N$, choose a random $r \in [0, {N}/{4}]$ and output the additive ciphertext as $E^+_{pk}(m) = \{{AC}_1, {AC}_2\}$, where ${AC}_1= (h^r\mod N)^N(1+mN)\mod N^2$; ${AC}_2=g^r\mod N$.

\textbf{Additive Decryption with Weak Private Key (AddDecWkey):} An additive ciphertext as $E^+_{pk}(m)$ can be decrypted with private key $sk = \theta$ by calculating:
$m=L\{\frac{{AC}_1}{[{({AC}_2)}^\theta\mod N]^N}\}$.

\textbf{Additive Decryption with Strong Private Key (AddDecSkey):} Use the strong key to decrypt $E^+_{pk}(m)$, $m$ is calculated as :$m = L[{({AC}_1)}^\lambda\mod N^2]\lambda^{-1}\mod N$, since $gcd(\lambda, N)=1$. The specific proofs are shown in Appendix A.

\textbf{Strong Private Key Splitting (SkeyS):}
According to the Chinese remainder theorem \cite{Chinese}, split the strong private key $\lambda$ into two partial strong private keys. One, denoted as $\lambda_i$, is sent to a party $i$; the other one, denoted as $\lambda_j$, is sent to a party $j$. $\lambda_i$ and $\lambda_j$ meet the following two constraints:

 $\begin{cases}
    \lambda_i + \lambda_j = 0\mod \lambda,\\
    \lambda_1 + \lambda_j = 1\mod N.
\end{cases}$

\textbf{Additive Decryption with Partial Strong Private Key Step One (AddDecPSkey1):}
The $U_i$ runs the  algorithm. Exploiting the partial strong private key  $\lambda_i$, the partial decrypted ciphertext ${DC}_1$ of $E^+_{{pk}}(m)$ can be calculated as:
${DC}_1 = ({AC}_{1})^{\lambda_i} = g^{\lambda_i \cdot r\theta}(1 + \lambda_imN)\mod N^2$. The  $U_i$ forwards $\{E^+_{{pk}}(m), {DC}_1\}$ to the  $U_j$.

\textbf{Additive Decryption with Partial Strong Private Key Step Two (AddDecPSkey2):}
This algorithm is run in the $U_j$'s side. Given $\{E^+_{{pk}}(m), {DC}_1\}$ and ${\lambda_j}$, the partial decrypted ciphertext ${DC}_2$ can be calculated as:
${DC}_2 = ({AC}_{1})^{\lambda_j} = g^{\lambda_j \cdot r\theta}(1 + \lambda_jmN)\mod N^2$. Then, the original message $m$ can be recovered as: $m=L[{DC}_1 \cdot {DC}_2]$. The specific proof is in Appendix B.

\textbf{Multiplicative Encryption (MulEnc):} Given a message $m \in {\mathbb{Z}}_N$, choose a random number $r \in [1, {N}/{4}]$ and output the multiplicative ciphertext as
$E^\times_{pk}(m) = \{{MC}_1, {MC}_2\}$, where $MC_1 = mg^{r\theta}\mod N$; ${MC}_2 = g^r\mod N$.

\textbf{Multiplicative Decryption (MulDec):} The multiplicative ciphertext $E^\times_{pk}(m)$ can be decrypted with the private key $sk=\theta$ by calculating :
$m = \frac{{MC}_1}{({MC}_2)^\theta}\mod N$.

\begin{figure*}[htp]
\centering
\includegraphics[scale=0.3]{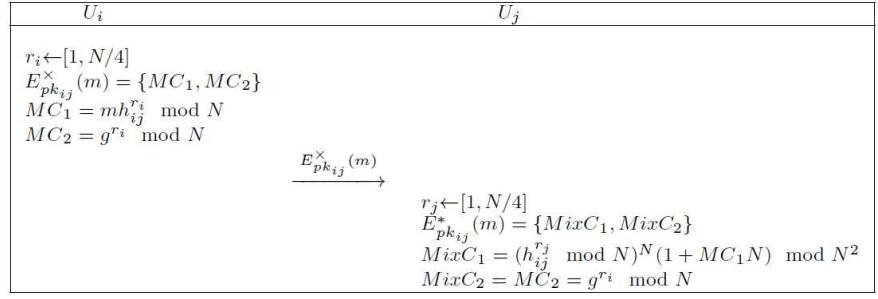}%
\caption{A Multiplicative Ciphertext to A Mixed Ciphertext (MultoMix).}
\label{MultoMix}
\end{figure*}

Seen from the \textbf{AddDecSkey} algorithm, our strong key $\lambda$ still can arbitrarily decrypt the additive ciphertext $E^+_{pk}(m)$. Thus, the additive ciphertext cannot be used directly.  We investigate the approach inspired by \cite{Lim}. Given a message $m$, we first run the \textbf{MulEnc} algorithm and acquire the multiplicative ciphertext $E^\times_{pk}(m)$, on which we next execute the \textbf{AddEnc} algorithm, and consequently we get a mixed cipheretex denoted as $E^{*}_{pk}(m)$, where $*$ indicates that a  multiplicative ciphertext is transformed to a mixed ciphertext. Consequently, an attacker, who gets $\lambda$, can only decrypt $E^{*}_{pk}(m)$and get $E^\times_{pk}(m)$, but not the original message $m$.  The joint key of the $U_i$ and the $U_j$ is $h_{ij}=g^{\theta_i\theta_j}\mod N$. This mixing process can be executed between  the $U_i$ and the $U_j$ as Fig. \ref{MultoMix}.  We introduce the \textbf{MultoMix} algorithm as follows.

\textbf{A Multiplicative Ciphertext to A Mixed Ciphertext (MultoMix):}

 \textbf{Step-1 (@$U_i$):} $U_i$ gives $U_j$ a multiplicative ciphertext $E^\times_{{pk}_{ij}}(m)=\{{MC}_{{ij},1}, C_{{ij},2}\}$, where ${MC}_{{ij},1} = mh_{ij}^{r}\mod N$; $C_{{ij},2} = g^r\mod N$.

  \textbf{Step-2 (@$U_j$):} $U_j$ randomly chooses a random $r'$, runs the \textbf{AddEnc} algorithm, and outputs the mixed ciphertext $E^*_{{pk}_{ij}}(m)$ as:
$E^*_{{pk}_{ij}}(m)=\{{MixC}_{{ij},1}, MixC_{{ij},2}\}$, where ${MixC}_{{ij},1}=(h^{r'}_{ij}\mod N)^N(1+mh_{ij}^{r}N)\mod N^2$; $MixC_{{ij},2}=C_{{ij},2}=g^r\mod N$.

Considering the mixed ciphertext $E^*_{{pk}}(m)$, if $mh_{ij}^{r}\mod N$ is denoted as $m'$, $E^*_{{pk}}(m)$ can be rewritten as :$E^*_{{pk}}(m)=\{{MixC}_{1}, MixC_{2}\}$, where ${MixC}_{1}=(h^{r'}_{ij}\mod N)^N(1+m'N)\mod N^2$; $MixC_{2}=C_{2}=g^r\mod N$. The mixed ciphertext actually is a variant of an additive ciphertext. If we use the generator $g$ of the  Paillier Cryptosystem, the message $mh_{ij}^{r}\mod N^2 \in Z_{N^2}$ is not bounded by $Z_N$. So the Paillier cryptosystem does not hold the \textbf{MultoMix} algorithm.

\begin{figure*}[htp]
\centering
\includegraphics[scale=0.3]{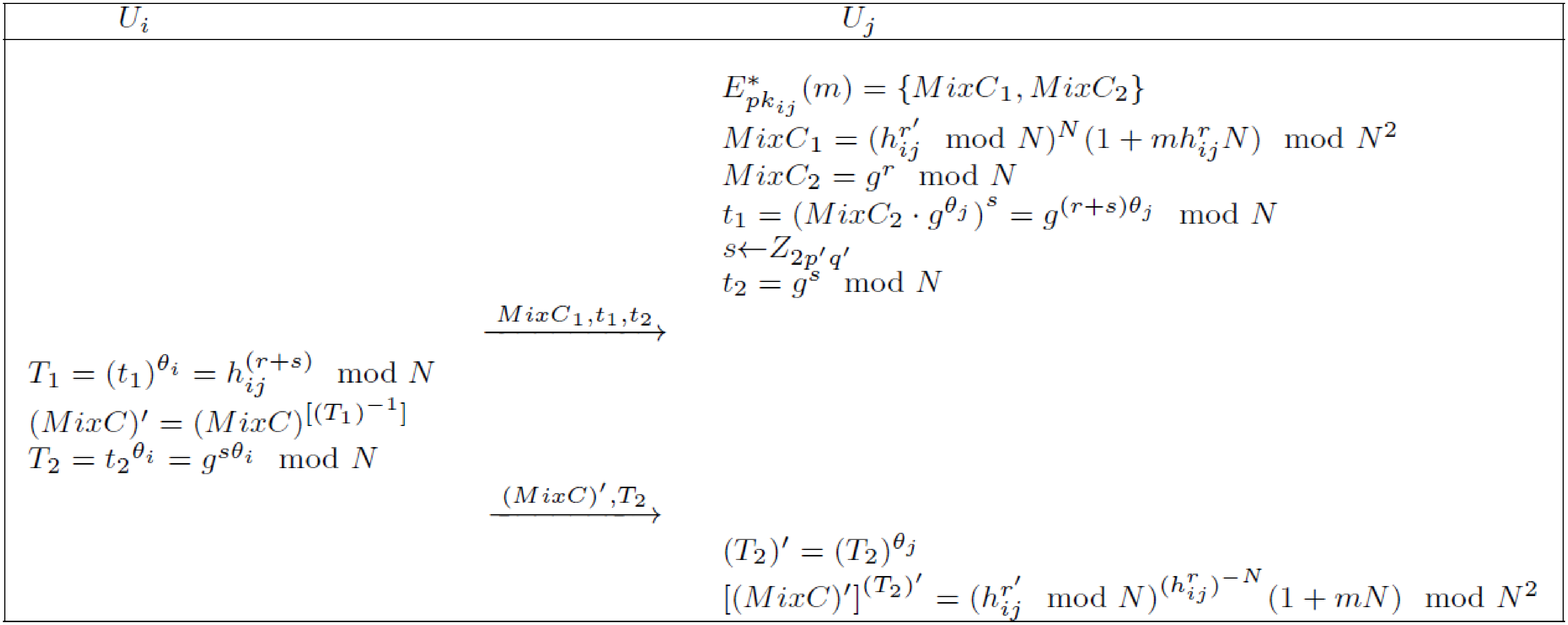}%
\caption{A Mixed Ciphertext to An Additive Ciphertext (MixtoAdd).}
\label{MixtoAdd}
\end{figure*}

To convert a mixed ciphertext to an additive ciphertext,  the process that a mixed ciphertext is transformed to an additive ciphertext is given as Fig. \ref{MixtoAdd}. We bring in the \textbf{MixtoAdd} algorithm  as follows.

\textbf{A Mixed Ciphertext to An Additive Ciphertext (MixtoAdd):}

 \textbf{Step-1 (@$U_j$):} Given a mixed ciphertext $E^*_{{pk}_{ij}}(m)=\{{MixC}_{{ij},1}, MixC_{{ij},2}\}$, where ${MixC}_{{ij},1}=(h^{r'}_{ij}\mod N)^N(1+mh_{ij}^{r}N)\mod N^2$; $MixC_{{ij},2}=g^r\mod N$, $U_j$ chooses a random number $s\in Z_{2p'q'}$, and computes $t_1={(MixC_{{ij},2} \cdot g^{\theta_j})}^s=g^{(r+s)\theta_j}\mod N$ and $t_2=g^s\mod N$. $U_j$ sends $\{{MixC}_{{ij},1}, t_1, t_2\}$ to $U_i$.

 \textbf{Step-2 (@$U_i$):} Once $t_1$  is received, $U_i$ first computes $(t_1)^{\theta_i}=g^{(r+s)\theta_i\theta_j}=h_{ij}^{(r+s)}\mod N$ and its inverse $[h_{ij}^{(r+s)}]^{-1}\mod N$, which $U_i$ uses to calculate $({MixC}_{ij})'$  as
follows:
\leftline{$({MixC}_{ij,1})'$}\\
\leftline{$=({MixC}_{ij,1})^{[h_{ij}^{(r+s)}]^{-1}}\mod N^2$}\\
\leftline{$=(h^{r'}_{ij}\mod N)^{{[h_{ij}^{(r+s)}]^{-1}}N}(1+mh_{ij}^{r}\cdot {[h_{ij}^{(r+s)}]^{-1}}N)\mod N^2$}\\
\leftline{$=(h^{r'}_{ij}\mod N)^{{[h_{ij}^{(r+s)}]^{-1}}N}[1+m{(h_{ij}^{s})}^{-1}N]\mod N^2$.}\\
 Then, $i$ computes $T_2=(t_2)^{\theta_i}=g^{s\theta_i}\mod N$, and forwards \{$({MixC}_{ij,1})', T_2$\} to $j$.

 \textbf{Step-3 (@$U_j$):} After receiving $T_2$, $U_j$ uses it to compute a middle result $(T_2)'=(T_2)^{\theta_j}=h_{ij}^s\mod N$. Next, the additive ciphertext $E^+_{{pk}_{ij}}(m)$ is computed as:
\leftline{$E^+_{{pk}_{ij}}(m)$}\\
\leftline{$=[({MixC}_{ij,1})']^{(T_2)'}$}\\
\leftline{$=(h^{r'}_{ij}~mod~N)^{{[h_{ij}^{(r+s)}]^{-1}\cdot h_{ij}^s}N}[1+m{(h_{ij}^{s})}^{-1}\cdot h_{ij}^sN]\mod N^2$}\\
\leftline{$=(h^{r'}_{ij}~mod~N)^{(h_{ij}^{r})^{-1}N}(1+mN)\mod N^2.$}\\

\section{A Simple Application of Restrained-Paillier Cryptosystem}
\begin{figure*}[htp]
\centering
\includegraphics[scale=0.3]{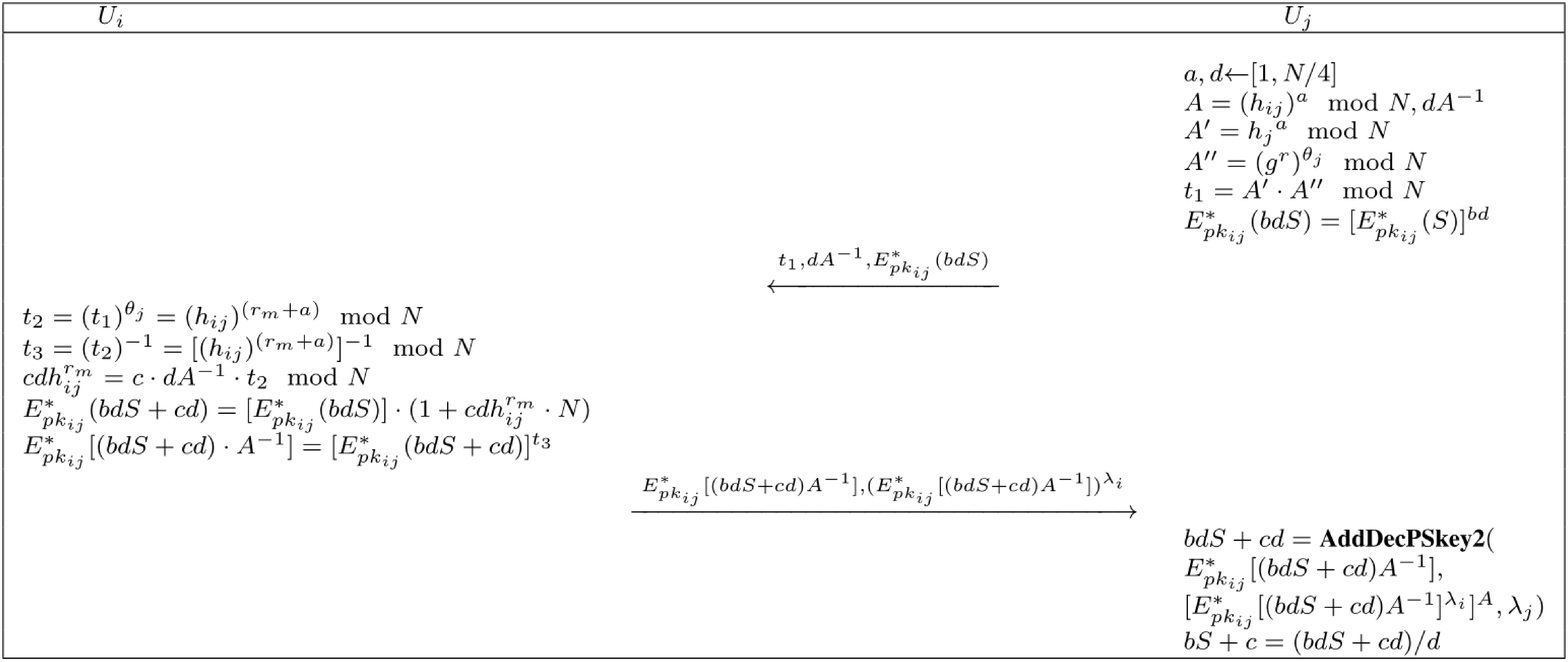}%
\caption{{ACCS} Protocol.}
\label{Access_Protocol}
\end{figure*}

$U_i$ and $U_j$ employ the Diffie-Hellman Key Exchange Agreement \cite{Diffie} to produce their joint public key $h_{ij}=g^{\theta_i\theta_j}\mod N$. The common secret $S$ between $U_i$ and $U_j$ is encrypted under the joint key $h_{ij}$, producing $E^*_{pk_{ij}}(S)=\{((h_{ij})^{r}\mod N)^N(1+S{(h_{ij})}^{r_m}N)\mod N^2; g^{r_m}\mod N\}$. $U_j$ obtains the secret $S$ on multiples of the number $b\in [1, N/8]$.  Besides, $U_i$ controls the access result using a control factor $c\in [1, N/8]$. $U_j$ gets the secret $bS + c$. The specific process seen in Fig. \ref{Access_Protocol}, is described as follows.
\subsection{\textbf{Access Control of Common Secret Protocol (ACCS)}}
\textbf{Step-1 (@$U_j$):} (1) The $U_j$ chooses $a$ and $d$ from $[1, \frac{N}{4}]$, and calculates ${A=(h_{ij})}^{a}\mod N$,  $dA^{-1}$ and $t_1=g^{\theta_j(r_m+a)}\mod N$. The parameter $d$ can hide $A^{-1}$.

(2) The $U_j$ calculates $E^*_{pk_{ij}}(bdS)=[E^*_{pk_{ij}}(S)]^{bd}$.  The $U_j$ keeps $A$ secret and sends $\{t_1, dA^{-1}, E^*_{pk_{ij}}(bd\cdot S)\}$ to the $U_i$.

\textbf{Step-2 (@$U_i$):} (1) The $U_j$ computes $t_2=(t_1)^{\theta_i}={(h_{ij})}^{(r_m+a)}\mod N$ and its inverse $t_3={(t_2)}^{-1}=[{(h_{ij})}^{(r_m+a)}]^{-1}\mod N$.

(2) The $U_j$ computes $cdh_{ij}^{r_m} = c\cdot dA^{-1}\cdot t_2$ and $E^*_{pk_{ij}}(bdS + cd)=[E^*_{pk_{ij}}(bdS)]\cdot(1+cdh_{ij}^{r_m}\cdot N)$.

(3) The $U_i$ computes $E^*_{pk_{ij}}[(bdS + cd)\cdot A^{-1}]=[E^*_{pk_{ij}}(bdS + cd)]^{t_3}$. For simplicity, $E^*_{pk_{ij}}[(bdS + cd)\cdot A^{-1}]$ is denoted as $Reslut_i$.

(4) The $U_i$ sends to the $U_j$ $\{Reslut_i, (Reslut_i)^{\lambda_i}\}$.

\textbf{Step-3 (@$U_j$):} (1) The $U_j$ obtains $bdS + cd$ as:
\leftline{$bdS + cd =\textbf{AddDecPSkey2}((Reslut_i)^{A}, [(Reslut_i)^{\lambda_i}]^{A},\lambda_j )$}.\\
(2) The $U_j$ obtains $bS + c = (bdS + cd)/d$.

\section{Three Protocols for Key Management and Identity Authentication}
The Paillier cryptosystem \cite{Paillier} is used for a single user because the Paillier only has a private key $\lambda$.
The modified Paillier cryptosystem in \cite{Bresson} and the proposed cryptosystem have the special characteristic that there is a strong private key and many weak keys. Because of the characteristic, the modified Paillier and the proposed cryptosystem can be used for multi-user systems. Weak private keys are individually distributed to users while the system manager has the only strong private key. In multi-user systems, key distribution and identity authentication are two important security issues.

We investigate the characteristic of the modified Paillier and the proposed cryptosystem to present three protocols for key management and identity authentication. We discuss the case where the KGC is a semi-trusted party and cannot know about each user's weak key. The weak key is processed by the it's user before sent to the KGC. When the user loses his/her weak key, he/she can take back the weak key from the KGC. We use one split part of the strong key to produce the identity certificates, and use the other split part to validate identity certificates. See the following protocols for details.

\subsection{\textbf{Identity Distribution and Key Management(IdDis \& KeyMan) Protocol}}
\begin{figure*}[htp]
\centering
\includegraphics[scale=0.3]{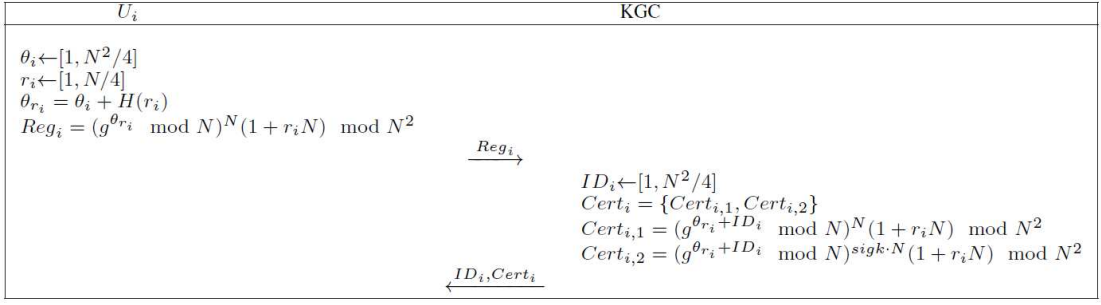}%
\caption{{IdDis \& KeyMan} Protocol.}
\label{fig_keyid}
\end{figure*}

The KGC has no access to private keys of parties. For example, the $U_i$'s private key is generated and hidden. Then, the $U_i$'s hidden key is regarded as the registration request, which is sent to the KGC. The KGC creates a unique identity ${ID}_i$, and uses it to generate an identity certificate ${Cert}_i$. The KGC returns $\{{ID}_i, {Cert}_i\}$  to $U_i$.  The KGC uses the \textbf{SkeyS} algorithm to produce a pair of split keys (${sigk}$, ${verk}$), where ${sigk}$ is used for signing ${Cert}$, and ${verk}$ is public for verifying ${Cert}$. When the $U_i$ enrolls in the system, the \textbf{IdDis\& KeyMan} protocol, seen as Fig. \ref{fig_keyid}, is implemented as follows.

\textbf{Step-1 (@$U_i$):} The $U_i$ chooses $\theta_i$ from [1, ${N^2}/{4}$] and $r_i$ from [1, ${N}/{4}$], computes $h_i=g^{\theta_i}\mod N$ as $U_i$'s public key, calculates $\theta_{r_i}=\theta_i+H(r_i)$ and preserves it as a secret. After that $U_i$ calculates ${Reg}_i=(g^{\theta_{r_i}}\mod N)^N (1+r_iN)\mod N^2$ and transmits it to the KGC as a registration request.

 \textbf{Step-2 (@$KGC$):} Once ${Reg}_i$ is received, the KGC assigns a unique identity ${ID}_i$ from [1, ${N^2}/{4}$] and produces a identity certificate ${Cert}_i$ by running the \textbf{AddDecPSkey1} algorithm as: ${Cert}_i= \{{Cert}_{i, 1}, {Cert}_{i, 2}\}$.\\
 \leftline{${Cert}_{i, 1} = (g^{{ID}_i}\mod N)^N{Reg}_i\mod N^2$}\\
 \leftline{$ \quad \quad \quad = (g^{\theta_{r_i}+{ID}_i}\mod N)^{N}(1+r_iN)\mod N^2$.}\\
 \leftline{${Cert}_{i, 2} = ({Cert}_{i, 1})^{{sigk}}\mod N^2$.}\\
The KGC stores $\{{ID}_i, {Cert}_i\}$ on its servers and returns $\{{ID}_i, {Cert}_i\}$ to the $U_i$.

\subsection{\textbf{Identity Authentication (IdAuth) Protocol}}
\begin{figure*}[htp]
\centering
\includegraphics[scale=0.3]{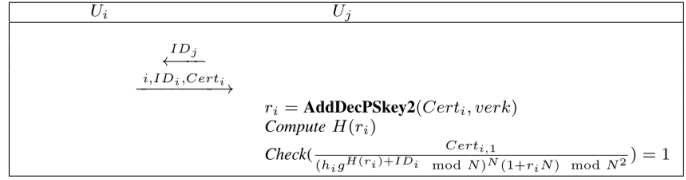}%
\caption{{IdAuth} Protocol.}
\label{IdAuth_protocol}
\end{figure*}

Before interaction, the $U_j$ verifies the $U_i$'s identity through running the \textbf{IdAuth} protocol, seen as
Fig. \ref{IdAuth_protocol}. This process is executed as follows.

\textbf{Step-1 (@$U_j$):} The $U_j$ sends the $U_i$ a request for identity authentication.

\textbf{Step-2 (@$U_i$):}  The $U_i$ returns \{$h_i$, ${ID}_i$, ${Cert}_i$\} to the $U_j$.

\textbf{Step-3 (@$U_j$):} The $U_j$ executes the \textbf{AddDecPSkey2} algorithm on ${Cert}_i$ using the verification key  ${verk}$, and obtains ${r}_{i}$. The $U_i$ uses it to compute $H(r_i)$, and computes $g^{H({r_i})}\mod N$. Then, $U_j$ checks whether the equation $\frac{{Cert}_{i,1}}{(h_ig^{H(r_i)+{ID_i}}\mod N)^N(1+{r}_{i}N)\mod N^2}$=1 holds. If it holds, the $U_j$ is safe in the knowledge of the $U_i$'s identity. Otherwise, the $U_j$ firmly refuses the $U_i$'s request.
\subsection{\textbf{Private Key Recovery(PriKeyRec) Protocol}}
\begin{figure*}[htp]
\centering
\includegraphics[scale=0.3]{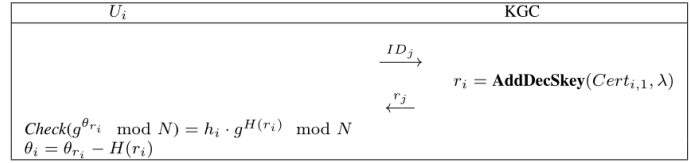}%
\caption{{PriKeyRec} Protocol.}
\label{Prikey_RecProtocol}
\end{figure*}

If the $U_i$ carelessly loses his/her own private key $\theta_i$, the $U_i$ can recover $\theta_i$ by running the \textbf{PriKeyRec} protocol, seen as Fig. \ref{Prikey_RecProtocol}. This process is described as follows.

\textbf{Step-1 (@$U_i$):} The $U_i$ issues a key recovery request ${ID}_i$ to the KGC.

\textbf{Step-2 (@$KGC$):} Once the key recovery request is received, the KGC runs the \textbf{AddDecSkey} algorithm on ${Cert}_{i,1}$ and gets a result, denoted as ${r}_{i}$ and sent to to the  $U_i$.

\textbf{Step-3 (@$i$):} The $U_i$ uses ${r}_{i}$ to compute $H(r_i)$, and computes $g^{H({r_i})}\mod N$. Then $U_i$  verifies whether the formula $g^{\theta_{r_i}} = h_i \cdot g^{H({r_i})}\mod N$ is correct. If not, the $U_i$ dispatches an error in calculation of $r_i$ to the KGC, and Step 2 and Step 3 are re-executed. Otherwise, the $U_i$ believes that ${r}_{i}$ is correct and takes out $\theta_i$ by calculating $\theta_i=\theta_{r_i}-H({r_i})$.

\section{Security Analysis}
In this section, we will analyze the security of the Restrained-Paillier, before demonstrating the security of four protocols. One protocol is used to achieving  the access control of common secret between two users. Other three protocols are used to realize  secure key management and identity authentication.
\subsection{Security of Restrained-Paillier}
Multiplicative ciphertexts of the Restrained-Paillier are secure under the INDistinguishable under Chosen
Plaintext Attack (IND-CPA) model since the multiplicative ciphertexts  same to those of the ElGamal scheme\cite{ElGamal}. The additive ciphertext of the Restrained-Paillier, formulated as $\{(h^r\mod N)^N(1+mN)\mod N^2; g^r\mod N\}$, has a N-power calculation more than those of the modified Paillier cryptosystem in \cite{Bresson}, formulated as $\{h^r(1 + mN)\mod N^2; g^r\mod N^2\}$. The security of additive ciphertexts of the Restrained-Paillier is still guaranteed. We give the secure analysis of our additive ciphertexts as follows.

\textbf{Against Additive Ciphertexts Analysis:}\\
The security of additive ciphertexts in the {Restrained-Paillier} can resist Chosen Plaintext Attack (IND-CPA) \cite{Katz} based on the Decisional Composite Residuosity (DCR) assumption in \cite{Paillier}. Let $N=pq$ be the product of two safe primes. The DCR assumption approximately claims that the set of \emph{N}-th powers modulo $N^2$ is computationally indistinguishable from the uniform distribution
modulo 	$N^2$, i.e.,

$Definition~1~(DCR~Assumption)$ £ºThe Decisional Composite Residuosity (DCR) assumption states that $\{x^N\mod N^2: x\in Z^\ast_{N^2}\}_{\approx}^{C}\{x:x\in Z^*_{N^2}\}$, where ${}_{\approx}^{C}$ denotes computational indistinguishability.

\emph{Theorem 1. If the DCR assumption holds, additive ciphertexts are secure under the INDistinguishable Chosen Plaintext Attack (IND-CPA)}.

\emph{Proof:}  Assume that a polynomial time distinguisher $\mathcal{A}$ chooses the messages $m_0$ and $m_1$ and sends them to a challenger $\mathcal{C}$. The public key is first set as ($N,g,h)$ where $h=g^\alpha\mod N$. $\mathcal{C}$ flips a random coin $\mu\in\{0,1\}$, randomly chooses \emph{r} and encrypts the data $m_\mu$, obtaining $\{c_\mu=[(g^{\alpha r}\mod N)]^N (1+m_\mu N)\mod N^2,g^r\mod N\}$ that is returned to $\mathcal{A}$.

In $\mathcal{A}$'s side, $c_\mu$ is the ciphertext of $m_0$ if and only if $\frac{c_\mu}{1+m_0N} = (g^{\alpha r}\mod N)^N\mod N^2$ is a \emph{N}-th residue as $g^{\alpha r}\mod N \in Z^*_N \in Z^*_{N^2}$. Therefore, a successful chosen-plaintext attacker could decide composite residuosity, and vice-versa.

\textbf{Against the \textbf{MultoMix} and \textbf{MixtoAdd} algorithms Analysis:}

The \textbf{MultoMix} and\textbf{ MixtoAdd} algorithms are secure under the IND-CPA model because the two algorithms are based on the $\textbf{AddEnc}$ and $\textbf{MulEnc}$ algorithms.

\textbf{Against Splitting of the String Key  Analysis:}

The privacy of divided private key is guaranteed by the Shamir secret sharing scheme \cite{Shamir}.  The strong private is randomly split into two shares in a way that any less than two shares cannot recover the original strong key. It further implies that the adversary cannot recover the original secret (the strong key) even the adversary gets the one share.

\subsection{Security of the ACCS Protocol}
The ciphertexts are secure under the IND-CPA model. The $U_i$ and $U_j$ can not use the extra data $<A, t_1, t_2, t_3>$ to decrypt ciphertexts, even though they have $\lambda$. $t_2$ is used to  present $U_i$ from computing $h_{ij}^{r_m}$.
Besides, $dA^{-1}$ is used to present $U_i$ from computing $h_{ij}^{r_m} = A^{-1}\cdot t_2\mod N$. Thus, the $U_i$, who has gotten $\lambda$, figures out $S(h_{ij})^{r_m}$ from the mixed ciphertext $E^*_{pk_{ij}}(S)$, but cannot decrypt $S(h_{ij})^{r_m}$  with $t_2=(h_{ij})^{(r_m+a)}$ or $dh_{ij}^{r_m}$.

\subsection{Security of  Three Key and Identity Protocols}
\emph{Theorem 2. For a passive adversary $\mathcal{A}$, the identity certificates in three protocols for key and identity are unforgeable}.

A passive adversary $\mathcal{A}$ gets the ${verk}$ that is public for verifying ${Cert}$, but cannot recover the the strong key since the Shamir secret sharing scheme is information-theoretic secure. Thus, $\mathcal{A}$ can not get the sign key ${sigk}$ and uses it to generate the signature of a message he chooses.

\emph{Theorem 3. For an active adversary $\mathcal{A}$, the identity certificates in three protocols for key and identity  are secure under Existential Unforgeability Against Adaptive Chosen Messages Attacks}.

The active adversary $\mathcal{A}$ gets ${Cert}_1$ and  ${Cert}_2$ from the certificate oracle. But $\mathcal{A}$ can not
produces a legitimate identity certificate that can pass validation, because of the properties of hash functions.

\emph{Proof:}{  }\\
\leftline{${Cert}_{1, 1} = (g^{\theta_1 + H({r_1}) + {ID}_1}\mod N)^{N}(1+r_1N)\mod N^2,$}\\
\leftline{${Cert}_{1, 2} = ({Cert}_{1, 1})^{sigk}.$}\\
\leftline{${Cert}_{2, 1} = (g^{\theta_2 + H(r_2)+ {ID}_2}\mod N)^{N}(1+r_2N)\mod N^2,$}\\
\leftline{${Cert}_{2, 2} = ({Cert}_{2, 1})^{sigk}.$}

$\mathcal{A}$ uses ${Cert}_1$ and  ${Cert}_2$ to generate ${Cert}_3$.
\leftline{${Cert}_{3, 1} = R^{N}(1+(r_1+r_2)N)\mod N^2,$}\\
\leftline{${Cert}_{3, 2} = ({Cert}_{3, 1})^{sigk},$}\\
where \leftline{$R= g^{\theta_1 + H({r_1}) + \theta_2 + H(r_2) + {ID}_1 + {ID}_2}\mod N$.} But ${Cert}_3$ can not pass validation, because of the properties of hash functions $H(r_1) + H(r_2) \neq H(r_1 + r_2)$.

\subsection{Analysis of the Security Requirements}
In this subsection, we show that the proposed protocol can achieve the design goals described in Section II.

\emph{Protection of Private Keys:} Participants' private keys are hidden in the registration requests $Reg$, identity certificates $Cert$, and public keys $h$.  $<Reg,  Cert, h>$ are multiplicative ciphertexts or additive ciphertexts which are secure under the IND-CPA model.

\emph{Correctness of Identity Certificates:} Correct identity certificate can pass the check in the \textbf{Auth} protocol.

\emph{Soundness of Identity Certificates:}  Use the corrupted identity certificate to get the result ${r'} = \textbf{AddDecPSkey2}({Cert}, {verk})$.  $H(r') \neq H(r)$ due to strong collision resistance fails to pass the check in the \textbf{Auth} protocol.

\emph{Unforgeability of Identity Certificates:}  The security requirement has been realized in \emph{Theorem 2}.

\emph{Existential Unforgeability Against Adaptive Chosen Messages Attacks:} The security requirement has been realized in \emph{Theorem 3}.

\emph{Recovery of Lost Private Keys:} The private key can be restored from the identity certificate. The corrupted identity certificate cannot pass the check in the \textbf{PriKeyRec} protocol.

\emph{Access Control of Common Secret:} Nobody can decrypt the encrypted secret by oneself. This has been proved in \emph{Security of the ACCS Protocol}.

\section{Performance Analysis}
In this section, we  analyze the communication cost and computation cost of the {Restrained-Pailiier} and our protocols.
\subsection{\textbf{Theoretical Analysis}}
We assume that one regular exponentiation operation with an exponent of length $|N|$ requires $1.5|N|$ multiplications \cite{Barker} (i.e. length of $r$ is $|N|$ and that computing $g^r$ requires $1.5|N|$ multiplications). Since the exponentiation operation brings in significantly higher cost than the addition and multiplication operations, we ignore the fixed numbers of addition and multiplication operations in our analysis.
\subsubsection{Restrained-Pailiier Cryptosystem} For comparison with the Paillier, the bit length of random number is chosen as $|r|=|N|/{4}$ and the bit length of private key is set as $|\theta|=|N^2|/{2}\approx |N|$. The \textbf{AddEnc} algorithm in the {Restrained-Pailiier} scheme needs $2.25|N|$ multiplications to encrypt a message, and the \textbf{AddDecWkey} algorithm consumes $3|N|$ multiplications to decrypt an aditive ciphertext while the \textbf{AddDecSkey} algorithm uses up $1.5|N|$ multiplications. The overhead comparison between the modified Paillier cryptosystem in \cite{Bresson} and Restrained-Paillier is listed in Table \ref{table_comparison}. The computation cost of the \textbf{AddEnc} algorithm in the {Restrained-Pailiier} is more $1.5|N|$ than that in the modified Paillier. It is because the calculation of $g^N$ in the {Restrained-Pailiier} occupies $1.5|N|$.
\begin{table}[!htbp]
\caption{Cost Comparison between Restrained-Paillier and Modified Paillier}
\label{table_comparison}
\centering
\begin{tabular}{|c|c|c|}
\hline
{}&${Restrained-Paillier}$&${Paillier}$\\
\hline
AddEnc &$2.25|N|$&$0.75|N|$ \\
\hline
AddDecSkey &$1.5|N|$&$1.5|N|$ \\
\hline
\end{tabular}
\end{table}

The \textbf{AddDecPSkey1} algorithm  and \textbf{AddDecPSkey2} algorithm both consume $3|N|$ multiplications, respectively as both the length of $\lambda_1$ and $\lambda_2$ is $2|N|$. The \textbf{MulEnc} algorithm occupies $0.75|N|$ multiplications, the \textbf{MulDec} algorithm needs $1.5|N|$ multiplications. The \textbf{MultoMix} algorithm requires $1.875|N|$ multiplications and the \textbf{MixtoAdd} algorithm needs $3|N|$ multiplications.

\subsubsection{Three Key Management and Identity Protocols}
The bit length of random number is chosen as $|r|=|N|/{4}$,  the bit length of private key is set as $|\theta|=|N^2|/{4}\approx |N|/2$, the bit length of identity is set as $|ID|=|N^2|/{4}\approx |N|/2$ and the bit length of hash function is set as $|N|/{4}$.

The computational costs of the three protocols are listed in Table \ref{table_four} with the KGC, $U_i$ and $U_j$ involved. For brief, we record the communication (Commu.) costs in last row of Table \ref{table_four}. In Table \ref{table_four}, the KGC's computation cost is $1.5|N|$ multiplications of the module-N's square, while the $U_i$'s computation cost is $1.5|N|$ multiplications of the module-N. Hence, the KGC uses more cost than  $U_i$ uses.
\begin{table}[!htbp]
\caption{ Costs of Three Key and Identity Protocols}
\label{table_four}
\centering
\begin{tabular}{|c|c|c|c|}
\hline
{}& {IdDis\&KeyMan} & {IdAuth} & PriKeyRec\\
\hline
KGC & $5.25|N|$ &  {} & $1.5|N|$ \\
\hline
 $U_i$	& $2.625|N|$  &{}& $1.5|N|$\\
\hline
$U_j$  &{}  & $5.625|N|$ &{}\\
\hline
Commu. &$6.5|N|$ &$6|N|$ &$0.75|N|$ \\
\hline
\end{tabular}
\end{table}

\subsubsection{ACCS Protocol}
The $U_j$ asks for the common secret with the help the $U_i$. The The computational (Compu.) cost of the ACCS protocol are listed in Table \ref{table_access}. The $U_j$ as the requester of common secret consumes more than the $U_i$, the helper. The communication cost is $8|N|$.
\begin{table}[!htbp]
\caption{ Costs of ACCS Protocol}
\label{table_access}
\centering
\begin{tabular}{|c|c|c|}
\hline
{}&${U_i}$&${U_j}$\\
\hline
Compu.&$5.625|N|$&$8.0625|N|$ \\
\hline
\end{tabular}
\end{table}
\subsection{\textbf{Experimental Analysis}}
We perform the experiments using a personal computer powered by an Intel(R) Core(TM) i5-4490 @3.30GHz processor, 8 GB of RAM memory and a Windows 7 professional operating system. The experimental results are averagely evaluated over 1000 times using a custom simulator built in Java.
\subsubsection{Restrained-Pailiier Cryptosystem}
The evaluation of the computational cost of the Restrained-Pailiier are listed in Table \ref{table_seven}. Computation cost  increases with increasing $|N|$. Seen from Table \ref{table_seven}, operations on multiplicative ciphertexts cost far less than operations on additive ciphertexts.  When $|N| = 1024$, time consumption is acceptable. But when $|N|=2048$,  time consumption is  too long.
\begin{table*}[thp]
\caption{The Performance of Restrained-Pailiier with Different $|N|/(ms)$}
\label{table_seven}
\centering
\begin{tabular}{|c|c|c|c|c|c|c|c|}
\hline
\diagbox{Algorithm}{$|N|$} & 512 & 768 & 1024 & 1280 & 1536	& 1792 &2048 \\
\hline
 AddEnc	& 1.398	& 4.511	& 9.875	& 19.743 & 32.903 & 50.88 & 75.209 \\
\hline
AddDecSkey & 1.108 & 3.84 & 8.529	& 17.284 & 28.81 & 44.89 & 66.343 \\
\hline
AddDecWkey & 1.664 & 5.27 & 11.397 & 22.55 & 37.325 & 57.7057 & 85.152\\
\hline
{AddDecSkey1} & 2.217 & 7.59 &	16.96 & 34.188	& 57.402	& 88.513 & 130.504\\
\hline
{AddDecSkey2} & 2.191 & 7.568 & 16.939 & 34.172 & 57.977 & 88.739 & 130.623\\
\hline
MulEnc & 0.296 & 0.679 & 1.397 & 2.612  & 4.324  &	6.792 & 9.767 \\
\hline
MulDec & 0.531 & 1.315 & 2.634 & 5.141 & 8.461	& 13.557 & 19.336 \\
\hline
\end{tabular}
\end{table*}
\subsubsection{Our Protocols}
\begin{figure*}[!thbp]
\centering
\subfloat[]{\includegraphics[width=3in]{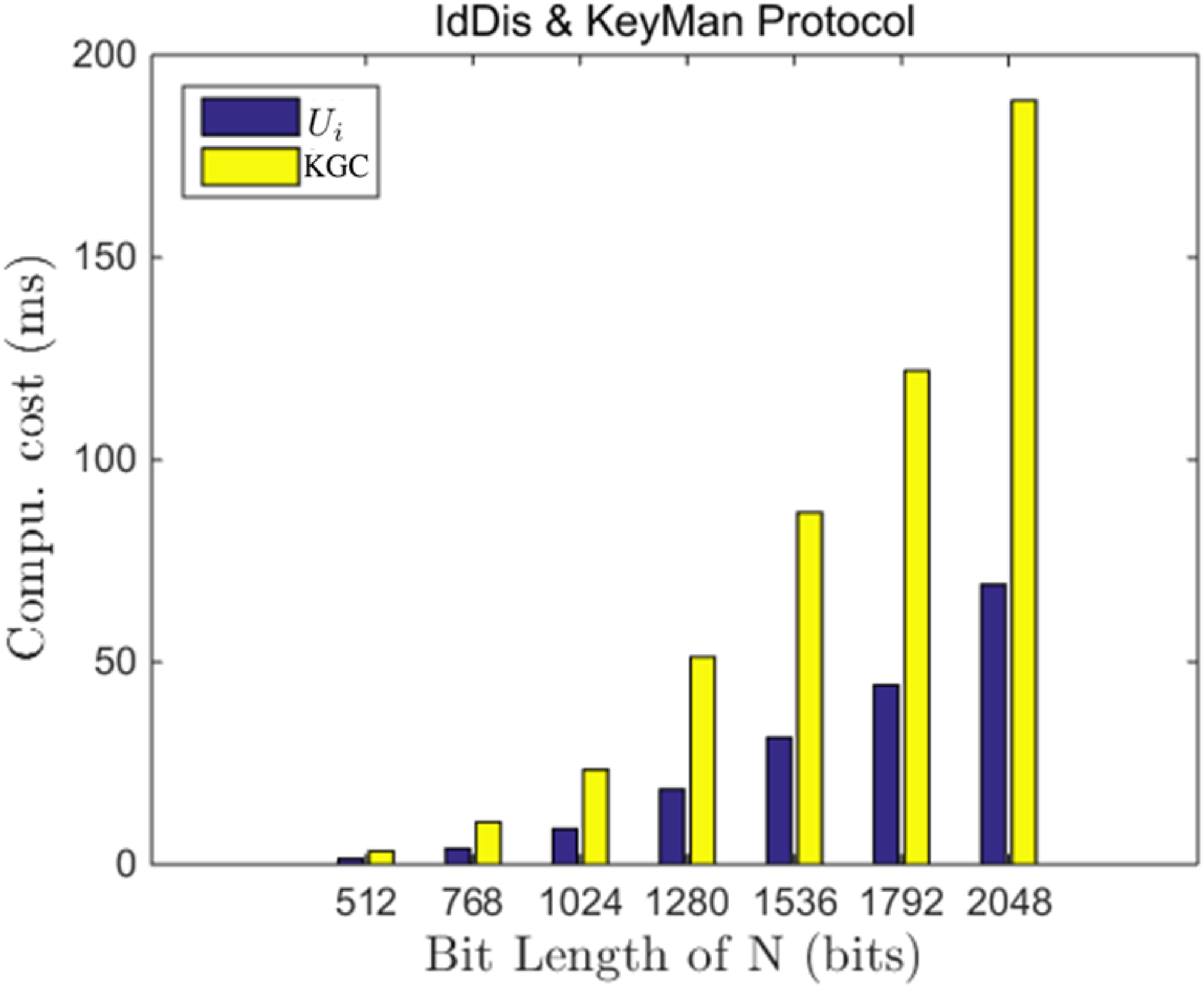}%
\label{idkey}}
\subfloat[]{\includegraphics[width=3in]{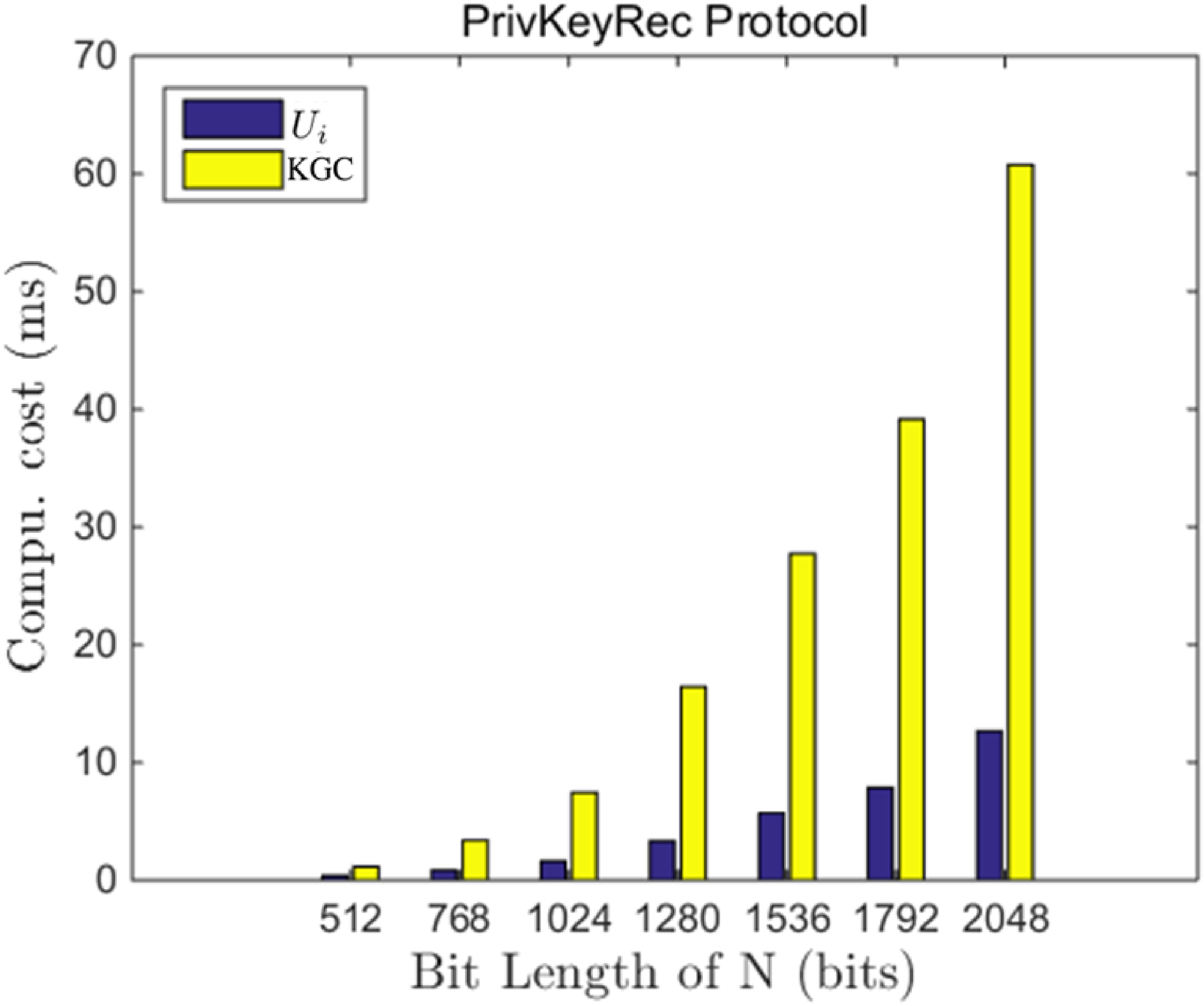}%
\label{reckey}}
\hfil
\subfloat[]{\includegraphics[width=3in]{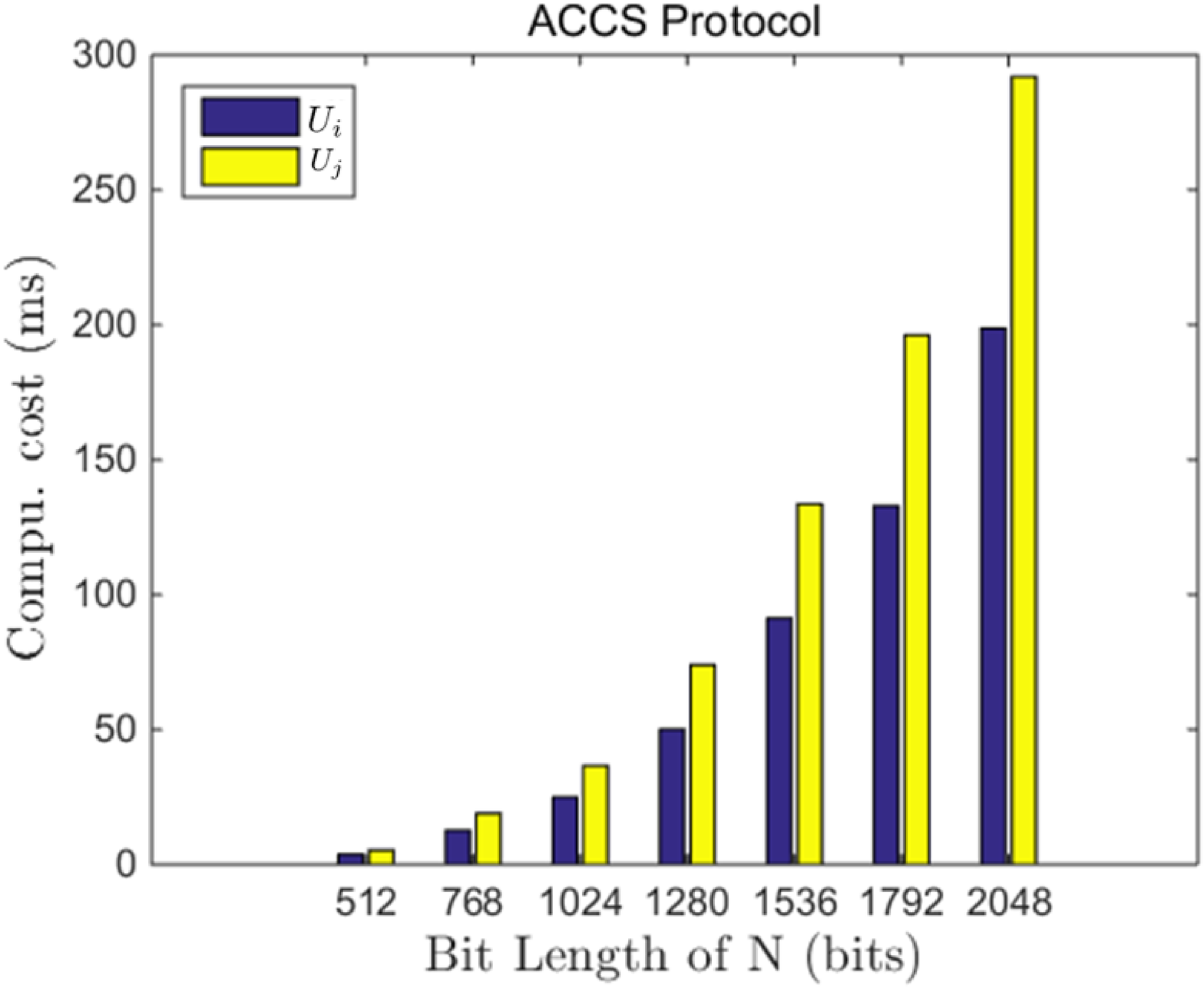}%
\label{auth}}
\subfloat[]{\includegraphics[width=3in]{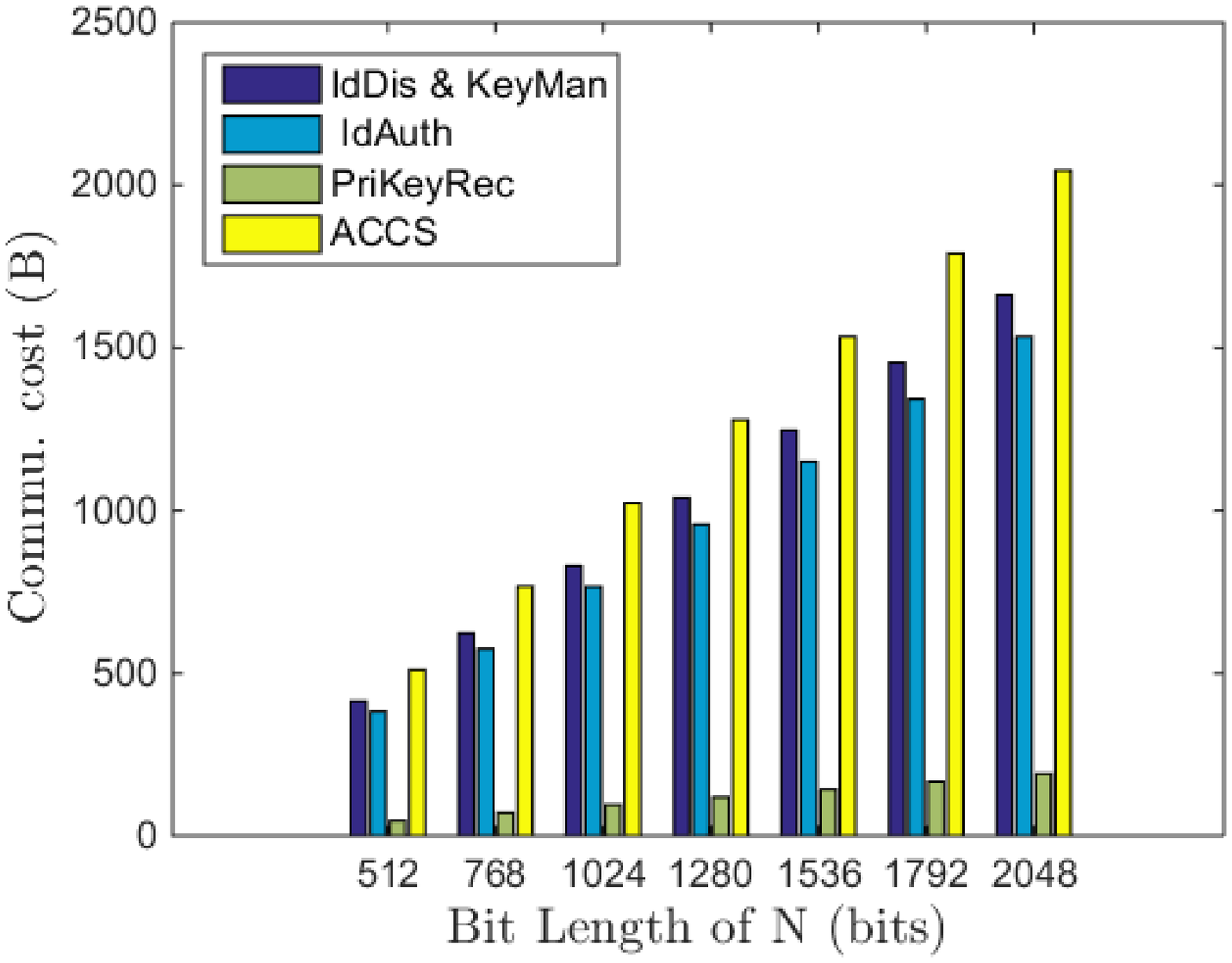}%
\label{commu}}
\caption{Evaluation findings. (a) Run time of the {IdDis \& Key Man} Protocol (vary with bit length of $N$). (b) Run time of the {PriKeyRec}  Protocol (vary with bit length of $N$). (c) Run time of the {ACCS} Protocol (vary with bit length of $N$).  (d) Communication cost of four protocols (vary with bit length of $N$).}
\label{fig_threeprotol}
\end{figure*}

The computational cost of \textbf{IdDis \& KeyMan} protocol is shown in Fig. \ref{fig_threeprotol}(a). The computational cost of \textbf{PrivKeyRec} protocol is shown in Fig. \ref{fig_threeprotol}(b). Seen from Fig. \ref{fig_threeprotol}(a) and Fig. \ref{fig_threeprotol}(b), the KGC uses more overhead than  $U_i$ uses.  The computation overhead of the \textbf{PrivKeyRec} protocol is far lower than that of the \textbf{IdDis \& KeyMan} protocol.
The computational cost of the \textbf{IdAuth} protocol is listed in Table \ref{table_auth}, where only the verifier needs to do calculations. The computational cost of the \textbf{ACCS} protocol, shown in Fig. \ref{fig_threeprotol}(c). The party requesting  the common secret $U_j$ costs more than the than the other party $U_i$ does.

\begin{table*}[!thbp]
\caption{The Performance of the IdAuth Protocol with Different $|N|$}
\label{table_auth}
\centering
\begin{tabular}{|c|c|c|c|c|c|c|c|}
\hline
\diagbox{IdAuth}{$|N|$} & 512 & 768 & 1024 & 1280 & 1536	& 1792 &2048 \\
\hline
 $U_j (ms)$	& 3.856	& 10.753 & 23.97& 51.637 & 87.415 & 122.662 &  190.095 \\
\hline
Commu. (B) & 382.413 & 575.19 & 766.523	& 958.19 & 1150.640 & 1342.426  &  1534.70 \\
\hline
\end{tabular}
\end{table*}

The communication overhead of the four protocol is shown in Fig. \ref{fig_threeprotol}(d), where $B$ represents 8 bits.  The communication overhead of the \textbf{ACCS} protocol is highest and the communication overhead of the \textbf{PrivKeyRec} protocol is lowest, seen from Fig. \ref{fig_threeprotol}(d).

\section{Conclusion}
We present the Restrained-Paillier cryptosystem, holding back the decryption of the strong key in the modified Paillier cryptosystem. The Restrained-Paillier provides both additive ciphertexts, multiplicative ciphertexts and mixed ciphertexts. Mixed ciphertexts can not be decrypted by the strong key. Using the the property of mixed ciphertext, we propose the ACCS protocol that can control the access to the common secret of two users. Moreover, exploiting the property of Restrained-Paillier, we put forward three protocols about the distribution of identity certificate, the identity authentication, key management and retrieval of private keys. Based on the security of Restrained-Paillier, the security of our protocols are guaranteed. We have calculated the computation and communication overheads of the Restrained-Paillier, our protocols. The simulation results show that these protocols use up less time consumption and low transmission cost. Hence, our protocols are attractive for practical applications. Future work will focus on how to apply the Restrained-Paillier onto digital signatures, especially proxy re-signatures.

\appendices
\section{}
Select a $a\in Z^*_{N^2}$ and  $a\mod N\neq1$. The Euler function is denoted as $\varphi(N)=(p-1)(q-1)=4p'q'$. Calculate the base $g$ as follows.
\begin{equation}
  \begin{split}
    g&=-{a}^{2N} \mod N\\
    &=-{(a\mod N)}^{2N}\mod N\\
    &=-{(a\mod N)}^{2pq\mod\varphi(N)}\mod N\\
     &=-{(a\mod N)}^{2(p-1)(q-1)}\mod N\\
    \end{split}
    \label{equa_one}
\end{equation}
For simplicity, let $\gamma={(a\mod N)}^{2(p-1)(q-1)}, \gamma\in{Z^*_N}$. The Equation (\ref{equa_one}) can be rewritten as the Equation (\ref{equa_two}).
\begin{equation}
 g=-{\gamma}^{2} \mod N\\
 \label{equa_two}
\end{equation}
Obtain the Equation (\ref{equa_three}) using the Equation (\ref{equa_two}).
\begin{equation}
 g^{\frac{\varphi(N)}{2}}=g^{\lambda}=1\mod N\\
 \label{equa_three}
\end{equation}

Since $g\in{Z^*_N}$, $g^{\theta r}\mod N \in{Z^*_N}$. Let $\mu= g^{\theta r}\mod N$. Since $N$ is the product of $p$ and $q$, $N$ has no primitive root. $\mu^{\frac{\varphi(N)}{2}}=\mu^{2p'q'}=\mu^\lambda =1\mod N$. Obtain $\mu^\lambda=1+kN\mod N^2 (0\leq k\leq N-1)$ and the Equation (\ref{equa_four}).
 \begin{equation}
   \begin{split}
    (\mu^\lambda)^N&=(1+kN)^N\\\
    &= 1+ C^1_N\cdot kN +\cdots+C^N_N\cdot(kN)^N\\
    &=1+kN^2\mod N^2\\
    &= 1\mod N^2
   \end{split}
   \label{equa_four}
 \end{equation}
 Thus, $({g^{\theta r}}\mod N)^\lambda\mod N^2=1\mod N^2$ holds.
\section{}
 In the \textbf{AddDecPSkey2} algorithm,  ${AC}^{(\lambda_1+\lambda_2)}=(g^{\theta r}\mod N)^{(\lambda_1+\lambda_2)N}(1+m(\lambda_1+\lambda_2)N)\mod N^2.$ We will prove that the method of splitting the strong key $\lambda$ can make \textbf{AddDecPSkey2} algorithm decrypt $m$.

In the splitting method, $\lambda_1+\lambda_2 \equiv n\lambda (0\leq n \leq \lambda-1)$ and $\lambda_1+\lambda_2 \equiv 1+kN (0\leq k \leq N)$. We can obtain the Equation (\ref{equa_seven}).
\begin{equation}
   \begin{split}
    {AC}^{(\lambda_1+\lambda_2)}&=(g^{\theta r}\mod N)^{n\lambda N}[1+m(1+kN)N]\\
    &= [1+m(1+kN)N]\mod N^2\\
    &=1+mN+mkN^2\mod N^2\\
    &= 1+mN\mod N^2
   \end{split}
    \label{equa_seven}
 \end{equation}

 The Equation (\ref{equa_seven}) can make $m=L[{AC}^{(\lambda_1+\lambda_2)}]$ hold.

\end{document}